\documentclass[12pt]{iopart}
\usepackage{graphicx}
\usepackage{color}

\begin{document}

\title{Impact of etches on thin-film single-crystal niobium resonators}

\author{H. Wang$^{1,4}$, T. Banerjee$^{2}$, T. G. Farinha$^{1,3}$, A. T. Hanbicki$^{1}$, V. Fatemi$^{2}$, B. S. Palmer$^{1,4}$, C. J. K. Richardson$^{1,3}$}

\address{$^1$ Laboratory for Physical Sciences, 8050 Greenmead DR, College Park, MD 20740\\
$^2$ School of Applied and Engineering Physics, Cornell University, Ithaca, NY 14853\\
$^3$ Department of Materials Science and Engineering, University of Maryland, College Park, MD 20742\\
$^4$ Department of Physics, University of Maryland, College Park, MD 20742\\
}

\ead{richardson@lps.umd.edu}
\vspace{10pt}
\begin{indented}
\item[]February 2024
\end{indented}

\begin{abstract}
A single crystal niobium thin film was grown using molecular beam epitaxy on a c-plane sapphire wafer. Several samples were fabricated into dc resistivity test devices and coplanar waveguide resonator chips using the same microfabrication procedures and solvent cleans. The samples were then subject to different acid cleaning treatments using different combinations of piranha, hydrofluoric acid, and buffered oxide etch solutions. The different samples expressed changes in dc resistivity in the normal and superconducting states such that the low temperature resistivities changed by more than 100\%, and the residual resistivity ratio dropped by a factor of 2. The internal quality factor of coplanar waveguide resonators measured near 5~GHz also showed significant variation at single photon powers ranging from 1.4$\times 10^6$ to less than 60$\times 10^3$. These changes correlate with the formation of surface crystallites that appear to be hydrocarbons. All observations are consistent with hydrogen diffusing into the niobium film at levels below the saturation threshold that is needed to observe niobium hydrides.
\end{abstract}

%

%
%
%
\ioptwocol

\section{Introduction}
The  elementary components of superconducting qubits are capacitors, inductors, and Josephson junctions. The composition of these various elements allow one to create different anharmonic resonant circuits, from a simple transmon, conceptualized as a single Josephson junction shunted by a capacitor, to more exotic circuits with more Josephson junctions \cite{siddiqi_engineering_2021, krantz_quantum_2019, kjaergaard_superconducting_2020}.

While the dc dissipation of the elements that make up the circuits is nominally zero, the rf or microwave dissipation is finite and ultimately limits the coherence of the device. For material sources of loss, the two leading candidates are dielectric and quasiparticle loss. In consideration of dielectric loss, all materials in proximity of the electric field can contribute. Superconducting quantum circuits are fabricated with careful selection of low-loss substrates such as sapphire and float-zone refined silicon, and thin film superconductors that form oxides and substrate interfaces that are low loss \cite{mcrae_materials_2023}. Fabrication hygiene is also important as residual resist from fabricating the circuits can limit the lifetime of the qubit \cite{richardson_materials_2020, de_leon_materials_2021}. Therefore, there is great interest in the materials and details associated with high-quality thin superconducting films with minimal oxide, defect, and interfacial losses and the processing of these materials into low-loss devices. 

For the superconducting metals used in qubits, a majority of devices have been historically fabricated using aluminum \cite{megrant_planar_2012}, a material that is relatively easy to deposit and forms a self-limiting oxide, which enables fabrication of all necessary circuit elements, including Al-AlO\textsubscript{x}-Al Josephson junctions. Nevertheless, a number of other superconducting materials have been investigated with respect to material loss, including the elemental metals niobium \cite{sage_study_2011} and tantalum \cite{barends_quasiparticle_2008, place_new_2021} and compounds such as refractory metal nitrides \cite{bruno_reducing_2015}. While the superconducting gap of these refractory metals is larger than Al, potentially reducing their sensitivity to pair-breaking radiation, focus has been on the metal's chemical robustness to strong acids and bases that enable the removal of trace chemical residues or control the relative ratios that different suboxide compounds form on the surface. While Al is most commonly found in the +3 oxidation state, the oxide formed as a byproduct of conventional nanofabrication processes results in a lossy film \cite{melville_comparison_2020} and the chemical reactivity of Al makes this oxide difficult to mitigate in typical academic nanofabrication facilities. Tantalum oxides are also primarily found with Ta in the +5 oxidation state \cite{place_new_2021, crowley_disentangling_2023}, but chemical processing can readily be used to remove and control the formation of tantalum oxides that exhibit lower loss \cite{crowley_disentangling_2023}. Similarly, Nb also forms multiple oxides in different oxidation states that have a variety of electrical properties; some of which exhibit high microwave loss \cite{cava_electrical_1991}. The impact of these various oxides on the superconducting loss of Nb thin film devices can also be mitigated by etching. Hydrofluoric acid has been used in etch protocols that improve superconducting Nb device performance \cite{alghadeer_surface_2023, altoe_localization_2022,murthy_developing_2022, murthy_tof-sims_2022, kowsari_fabrication_2021}.

Superconducting Nb has been studied extensively for use in radio frequency cavities that are a critical component for particle accelerators and a robust methodology has been established to minimize surface resistance \cite{kelly_surface_2017, myneni_medium_2023}. Some of these treatments include vacuum annealing to modify the native oxide composition \cite{ciovati_improved_2006}. Hydrogen incorporation is also a concern as it has been shown to be highly diffusive \cite{volkl_diffusion_1981}, can form defects \cite{cizek_hydrogen-induced_2004}, and modify the low-temperature electrical resistance \cite{isagawa_hydrogen_1980}. Interestingly, niobium pentoxide, Nb\textsubscript{2}O\textsubscript{5}, the low microwave loss phase of niobium oxide, is also desirable as a catalyst for hydrogen production and other acidic reactions \cite{zhou_recent_2020,zhao_nanostructured_2012}.

In this article, we report on microwave loss and dc measurements of a single crystal Nb (111) film after performing different acid cleans. The Nb film was deposited in ultra-high vacuum (UHV) using molecular beam epitaxy (MBE) \cite{wolf_epitaxial_1986} on a sapphire substrate and due to the growth conditions is single-crystal. To assess the loss of the film, a series of quarter-wave resonators from the Nb film were fabricated and measured at temperatures below 20 mK, which is a common characterization tool for loss \cite{mcrae_materials_2023}. While piranha cleans had a marginal improvement for the superconducting loss of the resonators, cleaning procedures that incorporate long times or concentrated HF increased the loss of the film, degraded the dc conductivity of fabricated strips, and even resulted in the formation of surface crystallites that form over time at room temperature. The surface morphology was measured with optical microscopy and atomic force microscopy (AFM), and chemical signatures were studied using x-ray photoelectron spectroscopy (XPS). 

\section{Thin film growth, processing, and cleaning}

\subsection{MBE growth}

Prior to growth, the backside of the wafer was coated with 50-nm of sputter deposited titanium to act as a heat spreader and ensure that the Nb layer was grown on a thermalized substrate. Prior to growth the wafer was degreased in a sequential series of beaker cleans of trichloroethylene, acetone (ACE), methanol, isopropyl alcohol (IPA), and deionized (DI) water and dried with a filtered nitrogen gun. The wafer is heated to 150 $^{\circ}$C for 30 min to desorb water and transferred to the UHV growth chamber. 

Single crystal superconducting Nb~(111) thin films were synthesized using a DCA M600 MBE system with an electron gun used to evaporate 99.99\% pure Nb metal onto a sapphire~(0001) wafer. Once inside the growth chamber, the wafer is soaked at 800~$^{\circ}$C for 15~min, followed by a heat spike to 1000~$^{\circ}$C to thermally desorb surface contaminants prior to growth. The 400-nm-thick Nb film is deposited at a substrate temperature of 800~$^{\circ}$C and growth rate of 0.02~nm/s. Following growth the substrate is heated to 1000~$^{\circ}$C and held for 1 hour. The substrate temperature was determined by the integrated thermocouple.

\begin{figure}[ht]
\begin{center}
\includegraphics[width=3.125in]{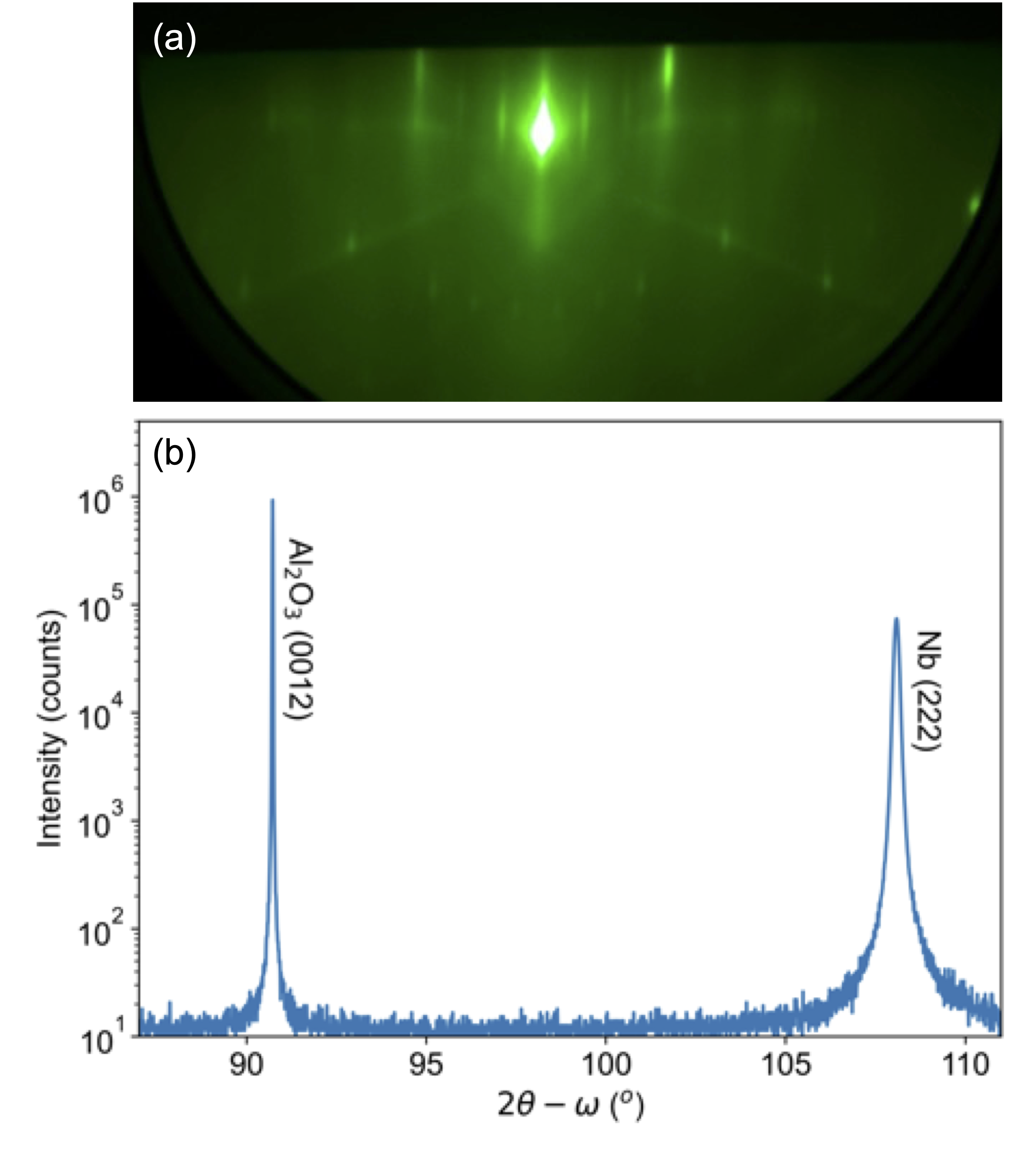}\\
\end{center}
\caption{(a) Post-growth, \textit{in-situ} reflection high energy electron diffraction pattern from the single crystal niobium thin film. (b) x-ray diffraction $\omega-2\theta$ measurement of the as-grown single crystal niobium film on sapphire.}
\label{RHEED-XRD}
\end{figure}

Reflection high energy electron diffraction (RHEED) was collected after the wafer was cooled to 200~$^{\circ}$C. The pattern in figure~\ref{RHEED-XRD}(a) shows a streaky 3x pattern indicating an ordered Nb surface. The symmetric x-ray diffraction measurement shown in figure~\ref{RHEED-XRD}(b) only shows peaks associated with sapphire \{0001\} and Nb \{111\} planes. The strong intensities indicate that the film has a high degree of uniformity and that Nb (111) is parallel to Al\textsubscript{2}O\textsubscript{3} (0001). The Nb (222) Peak at $2\theta-\omega=108.095^{\circ}$ is nearly completely relaxed with a residual biaxial in-plane strain of $\epsilon_x = -9.26\times 10^{-4}$. 

\begin{figure}
\begin{center}
\includegraphics[width=3.125in]{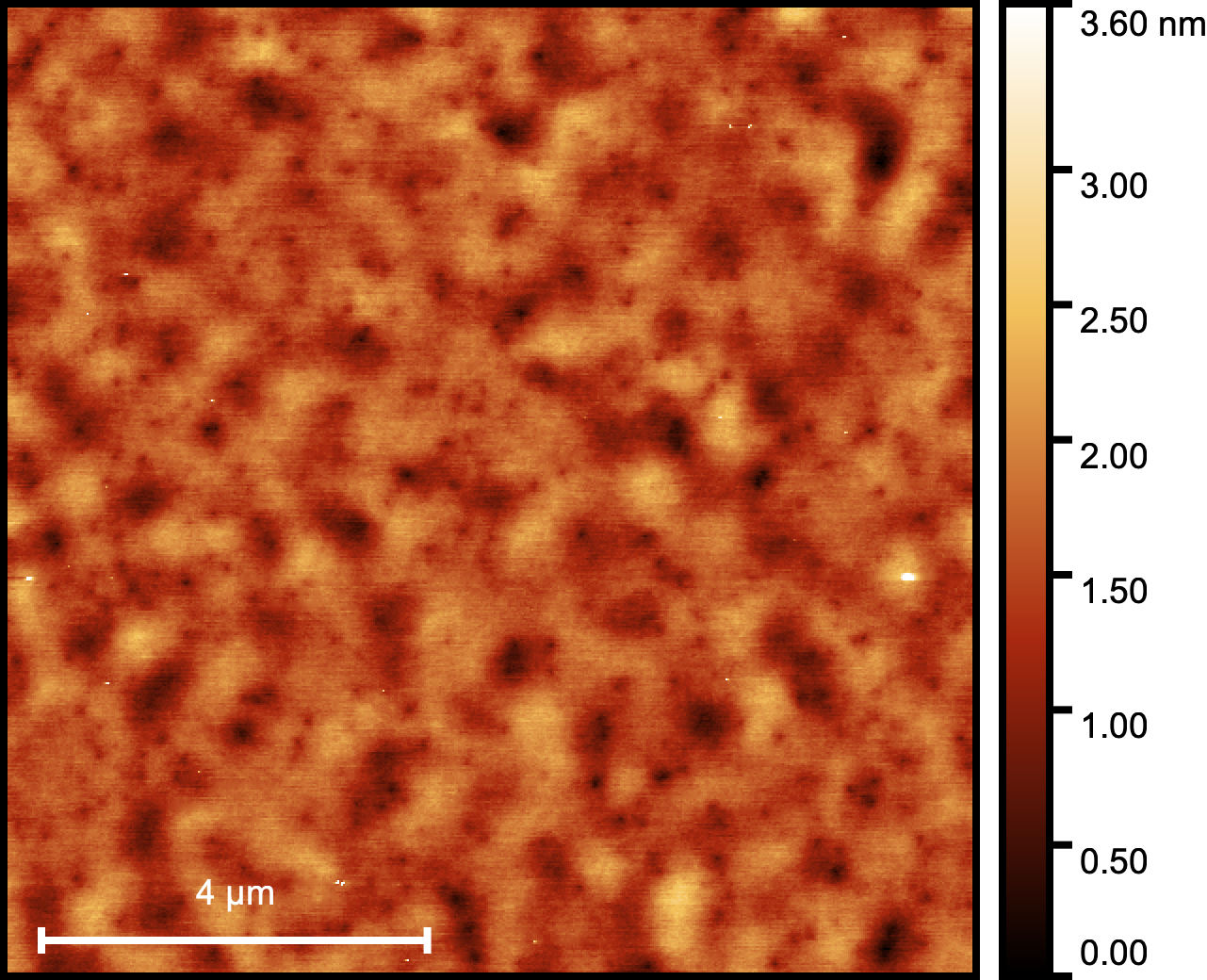}\\
\end{center}
\caption{A 10 $\times$ 10~{\textmu}m topographic atomic force microscope measurement of the niobium (111) surface before processing the film.}
\label{FilmAFM}
\end{figure}

The surface topography is measured by AFM and shown in figure~\ref{FilmAFM}. The 10 $\times$ 10 {\textmu}m scan exhibits a root mean square roughness of 328 pm, peak-to-peak roughness of 10 nm, and extrapolated autocorrelation length of 238 nm.  

\subsection{Device fabrication}
The backside Ti was removed using Transene Ti etchant TFT at 60~$^{\circ}$C before fabricating devices on the front side. To pattern the Nb film into devices for dc and resonator measurements, Olin OiR 908-35 positive photoresist is spun on the wafer to a nominal thickness of 3.5 {\textmu}m. Following pre-baking the sample, that pattern is exposed using a Heidelberg MLA150 direct laser writer. The sample undergoes a secondary post-bake, development using OPD4262, and rinsed in DI water. The patterned Nb sample is etched in an Oxford Cobra inductively coupled plasma reactive ion etcher (ICP-RIE) using an etch recipe consisting of a 2:3 mixture of Cl$_2$:BCl$_3$ with 350 watts power for 3 minutes for a complete etch through the entire film. 
    
Following patterning, the remaining photoresist is then removed by a 2-minute soak in ACE followed by a sequential rinse with a squirt bottle using ACE and IPA. The sample is further cleaned by soaking in a beaker of remover PG at ambient temperature for 12 hours, rinsed with DI water followed by filtered N$_2$ dry. To dice the sample into individual 5 $\times$ 5 mm die, a protective resist FSC-M is used on the sample. After dicing, the protective resist was removed by a 1-minute soak in a beaker with ACE, and then sequentially rinsed using ACE, IPA, and dried with filtered N$_2$.

\subsection{Acid cleans}
A series of fabricated devices were subjected to different acid treatments. The different acid cleaning protocols were all implemented after the device received identical solvent cleans as detailed above. In the procedures below, piranha is a mixture of sulfuric acid and hydrogen peroxide in a volume ratio of 3:1, which is commonly used to remove organic residues from surfaces of electronic devices. Buffered oxide etch (BOE), is an undiluted 6:1 volume ratio mixture of 40\% ammonium fluoride (NH$_4$F) in water to 49\% hydrogen fluoride (HF) in water. The hydrofluoric acid etchant is undiluted 49\% HF in water. All steps are beaker soaks with sufficient liquid to fully immerse the sample.
\begin{itemize}
    \item NC: No acid clean baseline/control
    \item P: Heated piranha etchant
    \begin{enumerate}
        \item 15 min heated piranha etch on hotplate set at 60~$^{\circ}$C
    \end{enumerate}
    \item PSB: Heated piranha + short BOE etchant 
    \begin{enumerate}
        \item 15 min heated piranha etch on hotplate set at 60~$^{\circ}$C
        \item 5 min undiluted, 6:1 buffered oxide etchant (BOE)
    \end{enumerate}
    \item PLB: Room temperature piranha + long BOE etchant
    \begin{enumerate}
        \item 15 min room temperature piranha etch
        \item 90 min, 6:1 buffered oxide etchant (BOE)
    \end{enumerate}
    \item HF: Room temperature HF etchant
    \begin{enumerate}
        \item 5 min room temperature undiluted 49\% hydrofluoric (HF) acid etch
    \end{enumerate}
\end{itemize}

While dilued solutoins are commonly used, undiluted etchants are used to accentuate any reactions with Nb. All treatment protocols were followed by a 1 min beaker rinse in DI water, a squirt bottle rinse with IPA, and blown dry using filtered nitrogen.

\subsection{Post-cleaning observations}
\begin{figure}[ht]
\begin{center}
\includegraphics[width=3.125in]{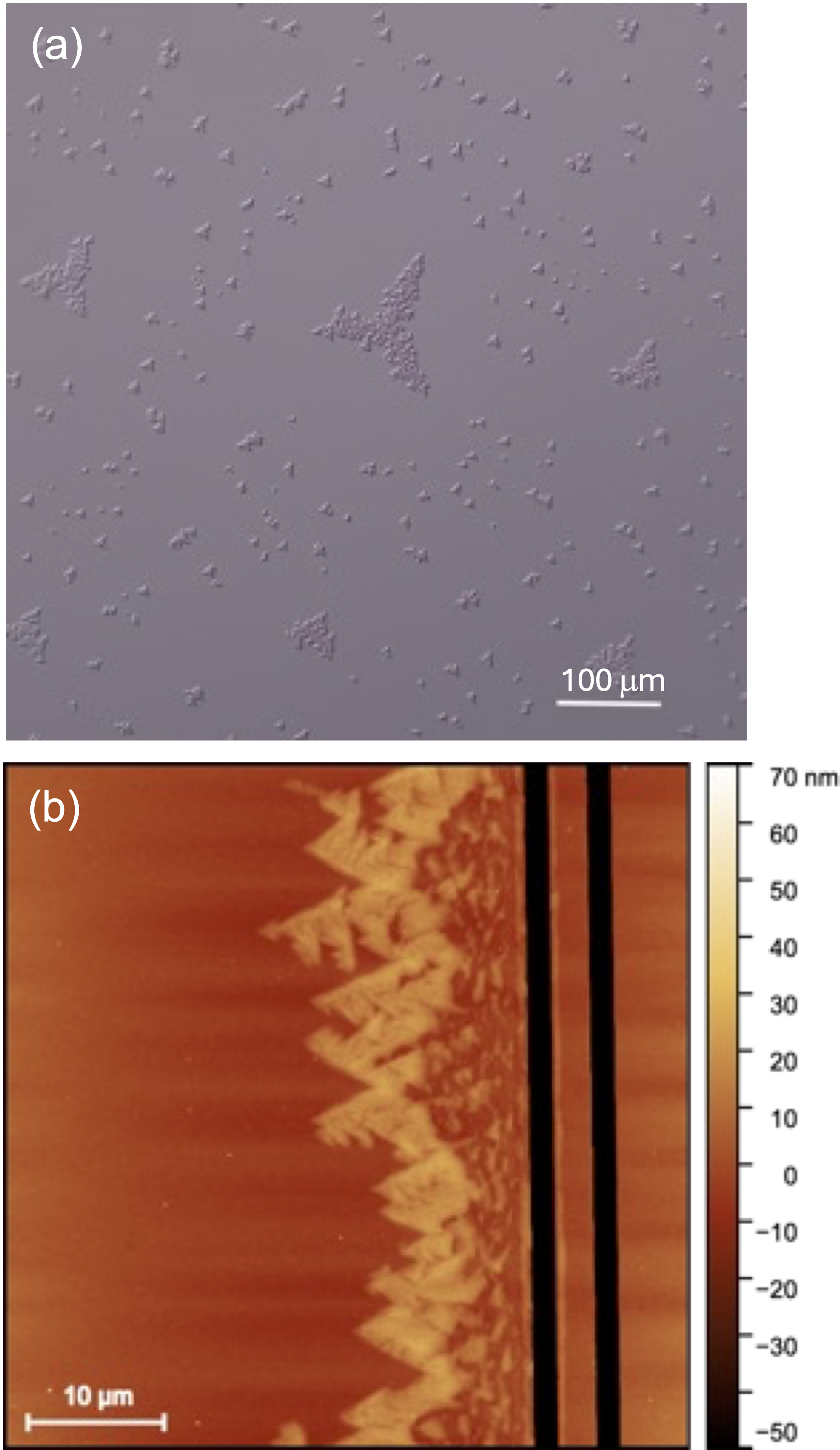}\\
\end{center}
\caption{Images of Nb surface defect regions. (a) A differential interference contrast  optical microscope image in the middle of the ground plane far from etched Nb features for the sample treated with just HF. (b) An atomic force microscope topography image near an etched resonator after PLB.}
\label{fig:NbDeformImage}
\end{figure}

Immediately following the cleaning treatments, each sample was inspected with an optical microscope. Every sample appeared clean with no detectable residue or defects. However, some samples showed defects that evolved while the samples were stored in a dry box, under N$_2$ purge. After approximately one day the HF and PLB samples showed extensive crystallites on the surface that decorated etched features and other scratches or pits on the devices. Typical images of defects are shown in the representative optical micrograph in figure~\ref{fig:NbDeformImage}(a) and AFM topograph in figure~\ref{fig:NbDeformImage}(b). The deformations are triangular and uniformly oriented across the sample. The 3-fold symmetric crystallite in the center of figure~\ref{fig:NbDeformImage}(a) has arms that are approximately 80-{\textmu}m long. Interestingly, each larger crystallite has a region around it that is depleted of smaller crystallites. This clear field is approximately 200 - 250 {\textmu}m in diameter. Confocal laser height (not shown) and AFM measurements indicated that these features are elevated mounds with a maximum height close to 50 nm above the smooth Nb surface. To further study these features x-ray photoelectron spectroscopic (XPS) measurements are performed.

\section{Chemical analysis}
\subsection{X-ray photoelectron spectroscopy}

The XPS measurements were done using a Thermo Nexsa G2 Surface Analysis system with a chamber pressure of $1.2\times 10^{-8}$ mBar. The system uses a monochromated Al K$\alpha$ x-ray source. Survey scans were conducted with an energy resolution of 0.4 eV and the individual peak scans had an energy resolution of 0.1 eV. The Fermi level is calibrated using a silver standard. Core level XPS data was collected as counts per second (CPS), and core level fittings were done using CasaXPS (version 2.3.25) using Voigt-like lineshapes and a Shirley background. All O 1s peaks are assumed to be Gaussian/Lorentzian mixed in equal ratios, GL(50). For the Nb 3d scan, the oxide and suboxide peaks were similarly fit using GL(50) lineshapes, while the metallic Nb were fit using Lorentzian asymmetric lineshape. The spin-orbit split for Nb was set to 2.72 eV, with the peak area ratio Nb3d$_{3/2}$:Nb3d$_{5/2}$ of 0.66:1 and full width at half maximum that are restricted to being identical for spin-orbit split peaks.

\begin{figure*}[ht]
\begin{center}
\includegraphics[width=\linewidth]{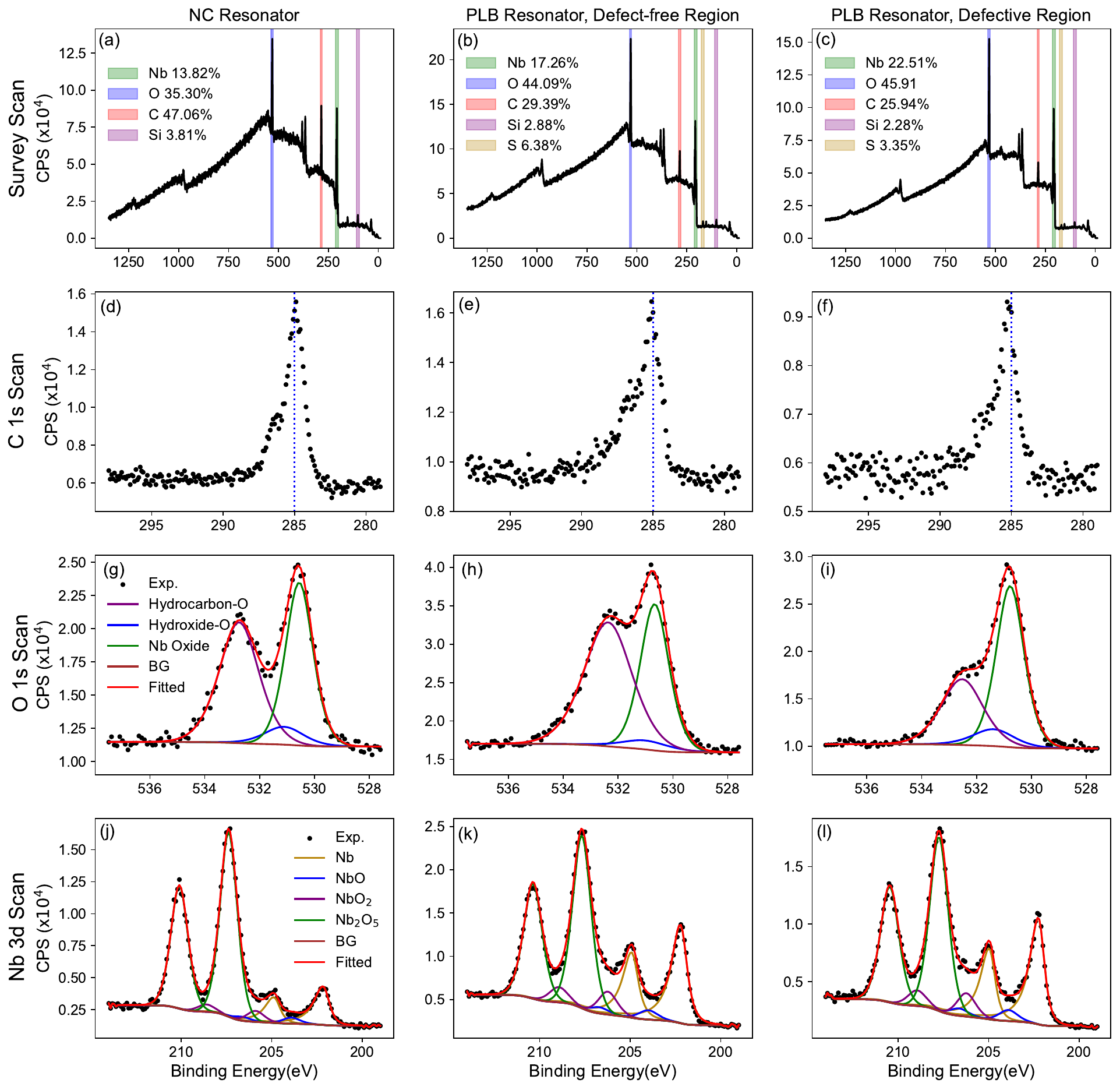}\\
\end{center}
\caption{X-ray photoelectron spectroscopic data for the three measured areas: the NC resonator (first column), the defect-free region of the PLB resonator (second column), and the defective region of the PLB resonator (third column). We include the survey scan ((a)-(c)), as well as the C 1s ((d)-(f)), O 1s ((g)-(i)), and the Nb 3d ((j)-(l)) scans for each area being scanned. Rows share the same legend.}
\label{fig:XPS_vert}
\end{figure*}

In order to examine the chemical differences between the defective and the defect-free regions of the PLB sample, XPS analysis of the two regions were conducted and compared to the results of the NC sample. The XPS scans are done to reveal the oxidation states of Nb as well as information regarding the oxygen and carbon present on the surface. Figure~\ref{fig:XPS_vert} shows the XPS data comparing the three regions using the survey scan as well as the C 1s, O 1s, and Nb 3d scans. The results of curve fitting contributions to the core-level spectra are given in table~\ref{table:XPS_Nb3d} for the Nb 3d peak and table~\ref{table:XPS_O1s} for the O 1s peak. 

\begin{table*}[ht]
\caption{X-ray photoelectron spectroscopic results of the core level fitting of the Nb 3d Data from the three measured regions. Nb 3d$_{5/2}$ binding energy (BE) and total atomic concentration At.\% is shown, Nb3d$_{3/2}$ BE is fixed at 2.72 eV above the respective Nb3d$_{5/2}$ BE.}
\footnotesize
\centering
\begin{tabular}{ccccccc}
\br
Peak&\multicolumn{2}{c}{NC Resonator}&\multicolumn{2}{c}{PLB, defect-free region}&\multicolumn{2}{c}{PLB, defect region}\\
 & d$_{5/2}$BE (eV) & At.\%& d$_{5/2}$BE (eV) & At.\%& d$_{5/2}$BE (eV) & At.\%\\
\mr
Nb  & 202.16 & 17.91\% & 202.22 & 33.90\% & 202.24 & 33.91\%\\
NbO & 203.92 & 2.47\% & 204.0 & 3.00\% & 203.91 & 3.49\%\\
NbO$_2$ & 205.89 & 4.49\% & 206.24 & 7.40\% & 206.23 & 7.41\%\\
Nb$_2$O$_5$ & 207.38 & 75.13\% & 207.67 & 55.71\% & 207.74 & 55.19\%\\
\mr
\end{tabular}
\label{table:XPS_Nb3d}
\end{table*}
\normalsize

\begin{table*}[h]
\caption{X-ray photoelectron spectroscopic results of the core level fitting of the O 1s Data from the three measured regions.}
\footnotesize
\centering
\begin{tabular}{ccccccc}
\br
Peak&\multicolumn{2}{c}{NC Resonator}&\multicolumn{2}{c}{PLB, defect-free region}&\multicolumn{2}{c}{PLB, defect region}\\
 & BE (eV) & At.\%& BE (eV) & At.\%& BE (eV) & At.\%\\
\mr
Nb Oxide    & 530.56 & 44.87\% & 530.67 & 39.67\% & 530.77& 56.64\%\\
Hydroxide   & 531.10 & 7.08\%  & 531.0 & 4.06\%   & 531.37 & 8.84\%\\
Hydrocarbon & 532.75 & 48.05\% & 532.36 & 56.27\% & 532.52& 34.52\%\\
\mr
\end{tabular}
\label{table:XPS_O1s}
\end{table*}
\normalsize

The survey scans reveal the presence of carbon, silicon, and sulfur contamination in both areas of the PLB sample (figure~\ref{fig:XPS_vert}(b),(c)), with the defect-free region containing a higher percentage of carbon and sulfur contamination. The C 1s scan shows a major peak at 285 eV (blue dotted line) from adventitious carbon. There is also an asymmetry towards higher binding energies, signifying the presence of oxygen-bonded carbon compounds. The Nb 3d scan for both regions (figure~\ref{fig:XPS_vert}(k),(l)) of the PLB sample are nearly identical, thus showing that there is no difference in the oxidation of the Nb in the two different regions. The Nb in both regions are primarily metallic Nb and Nb$_2$O$_5$, with small amounts of NbO and NbO$_2$. It is important to note that Nb hydrides cannot be reliably detected using XPS analysis, as it is well known that the XPS is unable to directly detect H. The Nb hydride peaks, if present, would be difficult to distinguish indirectly due to the presence of metallic Nb and Nb$^{2+}$.

The O 1s scan for both regions of the PLB sample shows two relatively distinct peaks (figure~\ref{fig:XPS_vert}(h),(i)). The peak at lower binding energy is attributed to niobium oxides. Given the low percentage of the the niobium suboxides, it would be difficult to fully separate the niobium oxides in the O 1s scan empirically, and thus are approximated using a single peak. Oxygen 1s peaks above 531 eV are generally attributed to oxygen from hydrocarbons \cite{beamson_appendices_1992}. The hydrocarbon peak also contains minor contributions from SiO$_2$ and possible sulphates that could have been formed on the surface after exposure to the pirahna etch. Owing to the significant overlap of the peaks, a unique fit is not feasible. Peak fitting was improved by including a contribution from surface hydroxides with binding energy close to 531~eV \cite{dupin_systematic_2000,frankcombe_interpretation_2023}. The hydrocarbon concentration is lower in the defect region than the defect-free region, which matches the relatively lower amount of carbon seen in the defect-free region. 

In comparison, the NC sample survey scan (figure~\ref{fig:XPS_vert}(a)), exhibits silicon and carbon contamination, but not sulfur contamination. Furthermore, the carbon contamination is significantly larger compared to either region of the PLB sample. The Nb is also more fully oxidized (figure~\ref{fig:XPS_vert}(j)), with lower proportions of suboxides and metallic Nb. The O 1s hydrocarbon peak is more separated and prominent compared to the niobium oxide peak. 

\section{Low temperature electrical characterization}
\label{Sec:Superconducting Measurements}

\subsection{DC resistance}
\begin{figure}[ht]
\begin{center}
\includegraphics[width=3.125in]{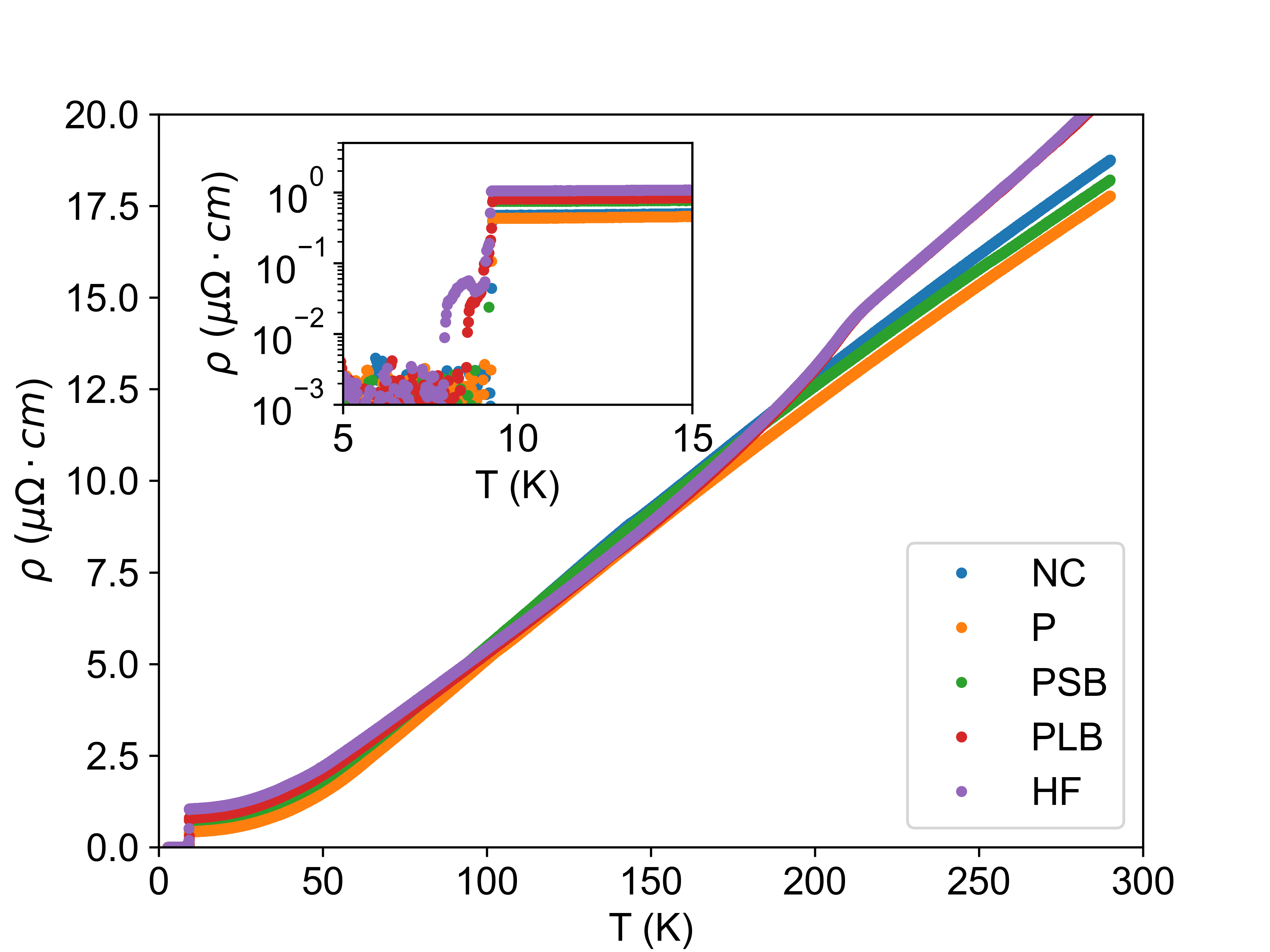}\\
\end{center}
\caption{Measured resistivity as a function of temperature while warming the niobium thin film samples following the different acid clean treatments.}
\label{fig:Tc}
\end{figure}

To measure the electrical characteristics of each sample, the dc resistivity at different temperatures is measured using a four-wire configuration. The samples are patterned using the Cl-based ICP-RIE etch described above into bar structures with two pairs of contact pads separated by a 100 $\times$ 5 {\textmu}m wire that is 420-nm thick. Each sample is individually mounted in a Sumitomo two-stage Gifford-McMahon cooler, and the current-voltage relationship is determined with a lock-in amplifier while the samples warm from 3.5 K to room temperature. A 13 Hz ac current of 10 {\textmu}A-rms was applied to the two outer contact pads of the bar structure and the voltage drop across the wire is measured between the two inner contact pads. The resistivity is calculated using Ohms law and the physical dimensions of the wire. The results for each sample are shown in figure~\ref{fig:Tc}. 

The observed transition between the normal and superconducting states initiates as a sharp transition for all samples, but there is a stepped transition between the normal and superconducting state observed for the PLB and HF samples. While the onset of the transition occurs at a uniform temperature, the complete transition is not abrupt for all samples. The reported value in table~\ref{tab:cleans} for $T_c$ is the temperature that the resistivity drops to the noise floor of the measurement. Results are listed in table~\ref{tab:cleans} for the room-temperature ($\rho_{RT}$) resistivity, room temperature fractional resistivity change between samples with the NC sample acting as reference ($\Delta\rho/\rho$), normal state low-temperature resistivity measured at $T\sim 10$ K ($\rho_{LT}$), residual resistivity ratio (RRR=$\rho_{RT}/\rho_{LT}$), and critical temperature ($T_c$). There is a clear and repeatable impact on the dc resistivity measurements of the various samples. The two samples that are not exposed to HF show the largest RRR and smallest resistances at both low temperatures and room temperature.

\begin{table*}[h]
\caption{Summary of dc resistance measurements for the different post-processing acid cleans: room-temperature ($\rho_{RT}$) resistivity, room temperature fractional resistivity change between samples with the NC sample acting as reference ($\Delta\rho/\rho$), normal state low-temperature resistivity measured at $T\sim 10$ K ($\rho_{LT}$), residual resistivity ratio (RRR), and critical temperature ($T_c$).}
\footnotesize
\centering
\begin{tabular}{cccccc}
\br
Treatment & $\rho_{RT}$ & ($\Delta\rho/\rho)_{RT}$ & $\rho_{LT}$ & RRR & $T_{c}$ \\
& ($\mu\Omega\cdot cm$) & & ($\mu\Omega\cdot cm$) & & (K) \\
\mr
NC & 18.7 & -- & .478 & 39.2 & 9.2 \\
P & 17.8 & -0.052 & .433 & 41.0 & 9.2 \\
PSB & 18.2 & -0.030 & .745 & 24.4 & 9.1 \\
PLB & 20.6 & 0.101 & .81 & 25.5 & 8.3 \\
HF & 20.7 & 0.104 & 1.05 & 19.7 & 7.9 \\
\br
\end{tabular}\\
\label{tab:cleans}
\end{table*}
\normalsize

\subsection{Microwave loss}
\label{sec:MicrowaveLoss}
The superconducting  loss is measured using coplanar waveguide (CPW) quarter-wave resonators coupled to a common transmission feed line. Each chip has 5 resonators, denoted \{R1, R2, R3, R4, R5\} with different lengths that determines their fundamental resonant frequencies $f_0$ that range from 5.3~GHz to 6.0~GHz. The resonators are designed with a coupling quality factor $Q_c$ of 250,000. To evaluate whether surface participation loss limits the devices \cite{sage_study_2011, mcrae_materials_2023}, the resonator's center trace width, w, is varied from 3 to 22 {\textmu}m. To maintain a 50 $\Omega$ characteristic impedance of the CPW, the gaps between the center trace and the ground plane, g, are appropriately varied from 1.5 to 11 {\textmu}m. 

Each resonator chip is glued to an Al sample box using GE varnish at the edges of the die \cite{huang_identification_2023}. Electrical connections from a surrounding printed circuit board to the chip are made with Al wirebonds. The packaged device is bolted to the mixing chamber of a Leiden cryogen-free dilution refrigerator. The input line has nominally: 20~dB attenuation on the still plate; 30~dB attenuation at the cold plate;  20~dB attenuator and low pass filter on the mixing chamber plate. The output line has a low pass filter followed by three isolators each contributing 20~dB of isolation on the mixing chamber plate and a high-electron-mobility-transistor (HEMT) amplifier on the 3-K plate (see Ref~\cite{yeh2017microwave} for further details of the setup). The complex transmission of the device under test is measured using a vector network analyzer (VNA) at different output source powers.

\begin{figure}[ht]
\begin{center}
\includegraphics[width=3.125in]{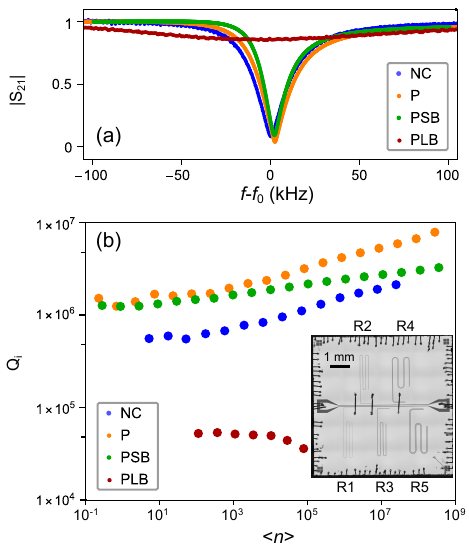}\\
\end{center}
\caption{(a) Resonator measurements near resonance transmitted magnitude, $|S_{21}(f-f_0)|$ of resonator R4 at relatively large power of internal photon number $\langle n \rangle \simeq 10^5$ for different acid cleaning treatments. (b) Extracted internal quality factors, $Q_i$, vs stored internal photon number, $\langle n \rangle$, for R4 for different acid cleaning treatments. The inset image is a stitched confocal laser image of the packaged resonator chip. The resonators R1 to R5 are arranged as indicated in the labels.}
\label{fig:resmeasure}
\end{figure}

Figure~\ref{fig:resmeasure}(a) show representative data of the measured transmitted data $|S_{21}(f-f_{0})|$ for resonator R4 for four of the five cleaned chips measured using a relatively large power of $\langle n \rangle = 10^5$ photons stored in the resonator on resonance. The PLB cleaned sample displays a significantly shallower and wider resonance compared with the other three cleaned chips, which is an indication that $Q_i$ is much smaller for that clean when compared with the three other measured data. Resonators from the HF device were not measured since the HF device displayed similar crystallites and degradation of its dc characteristics as the PLB device (see Table~\ref{tab:cleans}.
 
Quantitative assessment of each resonator is completed using the complex transmitted signal, $S_{21}(f)$ near each resonance and fitting the data using
\begin{equation}\label{eqn:DCM}
    S_{21}(f) = \alpha e^{(\phi+2\pi\tau f)}\left[1-\frac{Qe^{i\theta}}{Q_c\left(1+2 i Q \frac{f-f_0}{f_0}\right)}\right],
\end{equation}
where the complex prefactor $S_{\circ} = \alpha \exp(\phi+2\pi\tau f)$ is described using the normalization factor, $\alpha$, global phase, $\phi$, and uncorrected electronic path length compensation, $\tau$ \cite{GaoThesis}. The center frequency, $f_0$, and loaded quality factor, $Q$, are important fit results for each resonator. The internal quality factor is determined from the loaded quality factor, coupling quality factor, $Q_c$, and coupling phase, $\theta$ by $Q_i^{-1}=Q^{-1}-|Q_c^{-1}\cos{\theta}|$ \cite{khalil_analysis_2012}.

The  power applied to the device, $P_{\mathrm{app}}$, is determined using the output power of the VNA and the attenuation to the device measured at room temperature. The average photon number stored in the resonator \cite{bruno_reducing_2015},
\begin{equation}
\langle n \rangle = \frac{1}{\pi h f_{0}^2} \frac{Z_0}{Z_r} \frac{Q^2}{Q_c} P_{\mathrm{app}},
\end{equation}  
is determined using the characteristic impedance of the environment, $Z_0$ and the resonator, $Z_r$.

The power dependence of the internal quality factor is the inverse of the loss tangent, $Q^{-1}=\tan\delta$, and offers insight in to the microscopic mechanisms for loss. There are power independent losses from a variety of sources \cite{sage_study_2011}. The power dependent trend that is most relevant to this work is that of two level systems, which are known to act as saturable absorbers with well defined loss (internal quality factor) plateaus at high and low powers. Figure~\ref{fig:resmeasure}(b) shows a plot of the extracted $Q_i$ versus $\langle n \rangle$ for the R4 resonator. The results are also detailed in table~\ref{tab:Qi}.

\begin{table*} [h]
\caption{The nominal width of the center trace $w$, the center-ground gap size $g$, designed resonance frequencies and measured minimum $Q_i$ values measured in the low-power regime after different acid treatments of five different sized resonators mentioned in the main text. The coupling quality factor was designed to be 250,000 for each resonator. The measured $Q_c$ has an average value of 285,000 and a standard deviation of 78,000.}
\label{tab:Qi}
\centering
\footnotesize
\begin{tabular}{ccccccccc}
\br
& w & g & f$_0$ & \multicolumn{4}{c}{minimum power $Q_i (\times 10^6$)} \\
& ({\textmu}m) & ({\textmu}m) & (GHz) & NC & P & PSB & PLB\\
\mr
   R1 & 3 & 1.5 & 5.34 & 0.15  & 0.45  & 0.55   & 0.06  \\
   R2 & 6 & 3 & 5.40  & 0.31  & 0.56  & 0.67   & 0.09 \\
   R3 & 10 & 5 & 5.56 & 0.36  & 0.64  &  0.39  & 0.17 \\
   R4 & 16 & 8 & 5.73 & 0.56  & 1.52  & 1.27   & 0.05 \\
   R5 & 22 & 11 & 5.93 & 0.67  & 1.34  & 1.41   & 0.07 \\
\mr
\end{tabular}
\end{table*}
\normalsize

The baseline NC-chip, exhibits $Q_i \simeq 0.5 \times 10^6$ at low power that increases without saturation to $Q_i\sim 2 \times 10^6$ at the maximum power applied. The P-sample has a low-power $Q_i \sim 1.5 \times 10^6$ and a high-power $Q_i$ value around $8 \times 10^6$. At low powers, there is little difference between the P and PSB samples. At high powers, the PSB-sample has a $Q_i \sim 3 \times 10^6$ that is slightly lower than that of the P cleaned sample. The PLB-sample, with an exaggerated cleaning time in BOE results in a substantially decreased $Q_i \simeq5 \times 10^4$.  This sample also showed a weaker power-dependence. Similar behaviors are observed for $Q_i$ vs power for all of the resonators on the various chips.

\section{Discussion}

The methodologies applied in this study have not identified direct evidence linking hydrogen to the composition or electrical differences between the samples. Nor has the acid clean treatments been optimized to maximize the observed quality factor of Nb resonators. Instead, the context and consistency of the reported observations suggest that hydrogen incorporation is likely occurring during HF acid treatments. The single crystal nature of the studied film provides opportunities to observe these interactions in a different perspective than previous studies of polycrystalline Nb.

X-ray photoelectron spectroscopy analysis of the NC and PLB samples indicate that carbon contamination is reduced by the acid cleans. Additionally, the oxide thickness is reduced and the relative oxide compositions are comparable to that of other polycrystalline films \cite{altoe_localization_2022}. It should be noted that the oxides reported here are likely to be stable, as there was more than a year between film growth and the start of this work. The sulfur contamination observed in both regions of the PLB sample is due to the piranha acid clean that remains on the surface despite the extended HF treatment. It is important to note that the surface crystallites also form on the HF sample, suggesting that sulfur does not play a critical role in their formation. In addition to XPS analysis, this sample was measured using Raman spectroscopy, which has been shown to be sensitive to niobium hydrides with no indication of the anticipated well-defined peaks \cite{singh_raman_2016}.

The formation of surface crystallites that initially form near etched features and thin film defects suggests inhomogeneous nucleation. The uniform orientation of the triangular shapes indicate that diffusion is active and the underlying structure of the single crystal Nb film plays a role in crystallite nucleation and growth. It is likely that these crystallites have not been observed on polycrystalline Nb surfaces because of the non-uniformity introduced by grain boundaries and different crystal orientations. The crystallites on the surface of the PLB and HF samples that are reminiscent of snow flakes grow over the course of days. The formation of the larger crystallites and the clear fields surrounding them are consistent with a ripening process where the larger crystallites grow at the expense of the smaller ones. There is no evidence that these surface crystallites contain anything other than hydrogen, carbon, and oxygen. The necessary requirement of active diffusion is supported by the known bulk diffusivity of hydrogen \cite{volkl_diffusion_1981}, but the specific process is not clear because of the relative surface diffusivity of hydrocarbons is not quantified.

The impact of hydrogen on the normal state and superconducting properties of Nb are well studied \cite{halbritter_transport_2005,kelly_surface_2017}. The crystal structure, temperature, and dissolved hydrogen content of different samples must be considered when comparing electrical properties. For the previous work using polycrystalline foils and wires, the impact of microstructure is greater than the results reported here. Therefore, the room temperature fractional resistivity change at room temperature is compared instead of the low temperatures values. Using a cubic polynomial fit to the results of ref.~\cite{isagawa_hydrogen_1980}, the atomic concentration of hydrogen can be estimated. Based on the results in table~\ref{tab:cleans} and the assumption that the NC sample has a small but unquantified hydrogen background that is in thermodynamic equilibrium with the atmosphere, the observed reduction of the resistivity indicates that the P and PSB samples have an atomic hydrogen incorporation that is practically the same as the NC sample with an error that is $\pm$ 1 atomic \% H. Alternatively, the PLB and HF samples have an atomic hydrogen concentration that is approximately 2.2 atomic \%H higher than the NC sample.

This interpretation of the cause of the fractional change in resistivity is not an unreasonable assertion given the extensive evidence of hydrogen loading during electrochemical processes \cite{isagawa_hydrogen_1980} and the relatively low hydrogen suggested by our analysis. These estimated concentrations are significantly below the saturation concentration. The observations are consistent with hydrogen being completely dissolved in a single phase mixture with niobium, where the hydrogen occupies interstitial spaces in the BCC Nb lattice \cite{manchester_h-nb_2000}. The relatively low hydrogen levels also supports the lack of hydride observations in the chemical composition analysis that has been performed. Additionally, the change in behavior at the superconducting critical temperature, reduced RRR, and decreased quality factor are all consistent with hydrogen loading \cite{isagawa_hydrogen_1980,halbritter_transport_2005,kelly_surface_2017}. 

\section{Conclusions}

The single crystal niobium thin film grown on a c-plane sapphire wafer was fabricated into devices and subject to several acidic clean treatments that express differences in the dc resistivity in the normal and superconducting states, as well as significant reduction of the internal quality factor of coplanar waveguide resonators measured near 5~GHz. These changes correlate with the formation of surface crystallites that appear to be hydrocarbons. All observations are consistent with hydrogen diffusing into the niobium film at levels below the saturation threshold that is needed to observe niobium hydrides. Further studies are required to determine optimum etch procedures when using dilute HF solutions.

\section{Data availability statement}
The data that support the findings of this study are available from the corresponding author, CJKR, upon reasonable request

\section{Acknowledgements}

The authors thank Yi-Hsiang Huang for assembling the PCB used to perform the rf loss measurements.
The authors acknowledge the use of XPS facilities at Cornell supported by NSF through the Cornell University Materials Research Science and Engineering Center DMR-1719875.
V. F. is thankful for the support of the Aref and Manon Lahham Faculty Fellowship that contributed to engaging in this work. This material is based partially upon work supported by the Air Force Office of Scientific Research under award number FA9550-23-1-0706.

\section{References}
\bibliographystyle{iopart-num}
\bibliography{./NbCleaning,NbCleaning_bib_TB}

\end{document}